\documentclass[reprint,aps,superscriptaddress]{revtex4-2}
\pdfoutput=1
\usepackage{graphicx}
\usepackage{geometry}
\usepackage{multirow}
\usepackage{lipsum} 
\usepackage{tcolorbox}
\usepackage{bm}
\usepackage{xr}
\usepackage{float}
\usepackage{amsmath}
\usepackage{booktabs}
\usepackage{xcolor}
\usepackage{soul}
\usepackage{cleveref}
\usepackage{svg}
\usepackage[normalem]{ulem}
\usepackage{algorithm2e}
\usepackage{booktabs}
\usepackage{xspace}

\renewcommand{\sout}[1]{}
\SetKwComment{Comment}{/* }{ */}

\newcommand{\APS}{\textsc{aps}\xspace}
\newcommand{\IMDB}{\textsc{imdb}\xspace}
\newcommand{\HS}{\textsc{high school}\xspace}

\begin{document}

\title{Multiplex measures for higher-order networks}

\author{Quintino Francesco Lotito}
\email{quintino.lotito@unitn.it}
\affiliation{Department of Information Engineering and Computer Science, University of Trento, via Sommarive 9, 38123 Trento, Italy}

\author{Alberto Montresor}
\affiliation{Department of Information Engineering and Computer Science, University of Trento, via Sommarive 9, 38123 Trento, Italy}

\author{Federico Battiston}
\email{battistonf@ceu.edu}
\affiliation{Department of Network and Data Science, Central European University, 1100 Vienna, Austria}

\begin{abstract}
A wide variety of complex systems are characterized by interactions of different types involving varying numbers of units. Multiplex hypergraphs serve as a tool to describe such structures, capturing distinct types of higher-order interactions among a collection of units. In this work, we introduce a comprehensive set of measures to describe structural connectivity patterns in multiplex hypergraphs, considering scales from node and hyperedge levels to the system's mesoscale. We validate our measures with three real-world datasets: scientific co-authorship in physics, movie collaborations, and high school interactions. This validation reveals new collaboration patterns, identifies trends within and across movie subfields, and provides insights into daily interaction dynamics. Our framework aims to offer a more nuanced characterization of real-world systems marked by both multiplex and higher-order interactions.
\end{abstract}

\maketitle

\section{Introduction}
From biological organisms to social groups, both natural and artificial systems demand sophisticated modeling tools to accurately capture their fundamental properties. Understanding how to represent interactions in such complex systems is crucial for unraveling their intricate architecture and emergent functionality.
Networks have long offered a common language for studying these systems, representing units as nodes and interactions as dyadic links~\cite{boccaletti2006complex, cimini2019statistical}.
However, this approach overlooks group interactions involving three or more nodes, which are essential in systems where higher-order interactions are prevalent~\cite{battiston2020networks, battiston2021physics, torres2021why, battiston2022higher}. Examples include collaboration networks~\cite{patania2017shape}, human face-to-face interactions~\cite{cencetti2021temporal}, folksonomies~\cite{ghoshal2009random}, species interactions within complex ecosystems~\cite{grilli2017higher}, brain networks~\cite{petri2014homological}, and cognitive associations~\cite{citraro2023hypergraph}. 

Hypergraphs~\cite{berge1973graphs}, able to explicitly encode group interactions as hyperedges, have emerged as a popular framework to represent higher-order networks~\cite{battiston2020networks, battiston2021physics}.
Recently, significant effort has been dedicated to the characterization of hypergraphs, from centrality~\cite{benson2019three, tudisco2021node} and clustering~\cite{benson2018simplicial} measures, to backboning~\cite{musciotto2021detecting, musciotto2022identifying} and reconstruction~\cite{young2021hypergraph}. Investigations into higher-order networks cover both micro-~\cite{lotito2022higher, lotito2023exact, lee2020hypergraph} and mesoscale levels~\cite{wolf2016advantages,vazquez2009finding,carletti2021random,eriksson2021choosing, chodrow2021generative, contisciani2022inference, chodrow2023nonbacktracking,ruggeri2023community, lotito2023hyperlink}, unveiling structural principles essential to understanding group interactions in real-world systems. These higher-order interactions are known to impact
the dynamic and collective phenomena within networked systems~\cite{battiston2021physics}, affecting processes such as synchronization~\cite{skardal2020higher,millan2020explosive, lucas2020multiorder,gambuzza2021stability, zhang2023higher}, diffusion~\cite{schaub2020random,carletti2020random}, spreading~\cite{iacopini2019simplicial,chowdhary2021simplicial,neuhauser2020multibody} and evolution~\cite{alvarez2021evolutionary}.

However, not all interactions in complex systems are alike; they may differ in nature, type, and scope. This observation led researchers to introduce the concept of multilayer and multiplex networks~\cite{boccaletti2014structure, kivela2014multilayer}, where links are encoded into different interaction layers, each representing a distinct type of relationship~\cite{dedomenico2013mathematical, battiston2014structural}. Multilayer and multiplex networks can successfully describe systems such as trade networks~\cite{barigozzi2010multinetwork}, transportations networks~\cite{aleta2017multilayer}, collaboration networks~\cite{battiston2016emergence}, and the brain~\cite{de2017multilayer}.
Multiplex hypergraphs, where layers encoding hyperedges of different type, could offer a robust tool for describing complex systems that involve group interactions of varying types. Despite a few exceptions~\cite{vasilyeva2021multilayer} and a significant potential, however, multiplex hypergraphs remain relatively unexplored, and a general set of tools for their analysis is still missing. 

In this paper, we introduce some basic measures to characterize multiplex networks with higher-order interactions, spanning from the node/hyperedge level to the system's mesoscale. 
We propose measures for the activity of nodes in different layers and orders, as well as for node degree correlation. Moreover, we partition nodes in generalists or specialists based on how their higher-order degree is spread across layers. We characterize hyperedges by examining hyperedge order distributions and overlap in multiple layers. Additionally, we introduce measures to quantify the layer-dependent ability to connect either generalist or specialist nodes.  Finally, we study the correlation of community and core-periphery structures across layers. 

We apply our proposals to three different real-world datasets: scientific co-authorship in physics, collaborations in movies and face-to-face interactions in a high school. Our measures are able to highlight patterns of actors' collaborations across various film genres, discern co-authorship trends within and across physics subfields, and offer insights into the daily interaction dynamics of students. 

\section{Multiplex hypergraphs}
\begin{figure}
    \centering
    \includegraphics[scale=0.75]{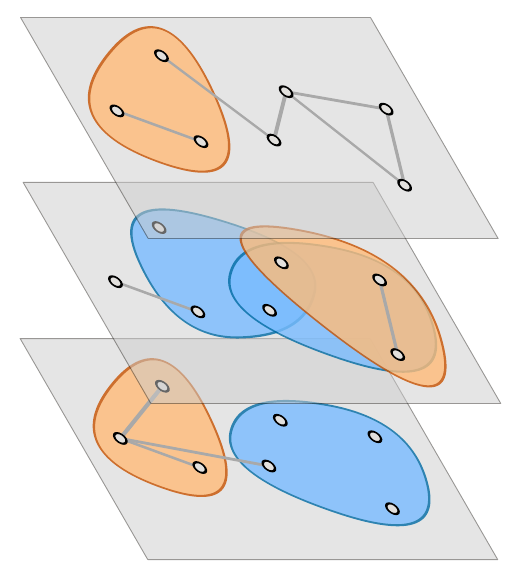}
    \caption{Multiplex hypergraphs represent systems of units that display interactions of different orders and different types. Each type of interaction is encoded into a single layer of the hypergraph. All the layers share the same set of nodes.}
    \label{fig:fig1}
\end{figure}

Multiplex hypergraphs model systems where interactions among units (\textit{i}) may belong to multiple types and (\textit{ii}) are not necessarily dyadic, i.e. they may involve more than two units. A \textit{multiplex hypergraph} $\textbf{H}$ is defined as:
\begin{equation*}
\textbf{H} = \{ H_1(V, E_1), \ldots, H_M(V, E_M)\}
\end{equation*}

where each \textit{layer} $\alpha$ is a hypergraph $H_{\alpha}(V, E_\alpha)$. Each hypergraph $H_{\alpha}(V, E_\alpha)$ share the same set of entities $V$. \( E_\alpha \subseteq \mathcal{P}(V) \) is the set of all interactions of a specific type $\alpha$. Moreover, we require \(|e| \geq 2\) for all \(e \in E_\alpha\) for any \(\alpha\). In other words, each layer in our framework shares the same set of nodes and represents a distinct set of interactions of the same nature. In Fig.~\ref{fig:fig1}, we show a simple multiplex hypergraph with $7$ nodes and hyperedges spread across $3$ layers.

We remark that our proposed framework is different from a multiplex representation of higher-order interactions where layers are defined by interactions of different order~\cite{lucas2020multiorder, sun2021higher}. 

In this work, we introduce a set of general tools to investigate multiplexity across different system scales in networks with higher-order interactions. We validate our measures and discuss relevant findings for three distinct real-world datasets: 
\begin{itemize}
\item \APS (Co-authorship network), where nodes are authors, and hyperedges represent groups of authors who have co-authored a paper. Each layer collects papers from the same subfield of physics, identified by a PACS code (Physics and Astronomy Classification Scheme)~\cite{pacs_data}.
\item \IMDB (Co-starring network), where nodes represent actors, and hyperedges represent the cast of a specific movie. Each layer corresponds to a movie genre.
\item \HS (Social network), where nodes are students, and hyperedges represent groups of students interacting face-to-face, with each layer grouping interactions from the same day of the week~\cite{mastrandrea2015contact}.
\end{itemize}
Detailed statistics about datasets are provided in Table~\ref{tab:datasets}.

\begin{table*}[ht]
\centering
\resizebox{0.75\textwidth}{!}{%
\begin{tabular}{cccccccccccc}
\hline
\multicolumn{4}{c}{\APS}                                              & \multicolumn{4}{c}{\IMDB}                                            & \multicolumn{4}{c}{\HS}           \\ \hline
Layer                           & $N$      & $E$      & \multicolumn{1}{c|}{$\bar d$}    & Layer                           & $N$      & $E$     & \multicolumn{1}{c|}{$\bar d$}    & Layer                           & $N$   & $E$    & $\bar d$   \\ \hline
\multicolumn{1}{c|}{AMPhys}     & 30375  & 12562  & \multicolumn{1}{c|}{3.7}  & \multicolumn{1}{c|}{Anim}  & 5545   & 864   & \multicolumn{1}{c|}{9.5} & \multicolumn{1}{c|}{Mo}         & 312 & 2655 & 1.3 \\
\multicolumn{1}{c|}{CM1}        & 63919  & 27241  & \multicolumn{1}{c|}{3.2}  & \multicolumn{1}{c|}{Comedy}     & 69303  & 9111  & \multicolumn{1}{c|}{13.6} & \multicolumn{1}{c|}{Tu}         & 310 & 3002 & 1.2 \\
\multicolumn{1}{c|}{CM2}        & 103075 & 58075  & \multicolumn{1}{c|}{4.0}  & \multicolumn{1}{c|}{Doc}        & 13357  & 2007  & \multicolumn{1}{c|}{7.4}  & \multicolumn{1}{c|}{We}         & 303 & 2543 & 1.2 \\
\multicolumn{1}{c|}{EMag}       & 57056  & 30908  & \multicolumn{1}{c|}{2.8}  & \multicolumn{1}{c|}{Drama}      & 103163 & 15384 & \multicolumn{1}{c|}{12.6} & \multicolumn{1}{c|}{Th}         & 295 & 2529 & 1.2 \\
\multicolumn{1}{c|}{EPart}      & 62997  & 26703  & \multicolumn{1}{c|}{62.2} & \multicolumn{1}{c|}{Family}     & 12968  & 1274  & \multicolumn{1}{c|}{12.6} & \multicolumn{1}{c|}{Fr}         & 299 & 2339 & 1.2 \\
\multicolumn{1}{c|}{GAA}        & 41535  & 10670  & \multicolumn{1}{c|}{9.6} & \multicolumn{1}{c|}{Fantasy}    & 14793  & 1136  & \multicolumn{1}{c|}{15.0} & \multicolumn{1}{c|}{}           &     &      &     \\
\multicolumn{1}{c|}{GasPhy}     & 16182  & 5120   & \multicolumn{1}{c|}{5.3}  & \multicolumn{1}{c|}{Horror}     & 28254  & 2964  & \multicolumn{1}{c|}{11.6} & \multicolumn{1}{c|}{}           &     &      &     \\
\multicolumn{1}{c|}{Gen}        & 69074  & 35940  & \multicolumn{1}{c|}{3.2}  & \multicolumn{1}{c|}{Thriller}   & 44188  & 4739  & \multicolumn{1}{c|}{13.8} & \multicolumn{1}{c|}{}           &     &      &     \\
\multicolumn{1}{c|}{IntPhy}     & 48136  & 17382  & \multicolumn{1}{c|}{3.0}  & \multicolumn{1}{c|}{}           &        &       & \multicolumn{1}{c|}{}     & \multicolumn{1}{c|}{}           &     &      &     \\
\multicolumn{1}{c|}{NPhy}       & 50142  & 20672  & \multicolumn{1}{c|}{16.7} & \multicolumn{1}{c|}{}           &        &       & \multicolumn{1}{c|}{}     & \multicolumn{1}{c|}{}           &     &      &     \\ \hline
\multicolumn{1}{c|}{\textbf{H}} & 315421 & 219769 & \multicolumn{1}{c|}{12.2} & \multicolumn{1}{c|}{\textbf{H}} & 195377 & 37465 & \multicolumn{1}{c|}{12.6} & \multicolumn{1}{c|}{\textbf{H}} & 327 & 7818 & 1.3
\end{tabular}%
}
\caption{Statistics about three real-world multiplex hypergraphs. For each layer, we report the number of active nodes $N$, the number of hyperedges $E$ and their average order $\bar d$. In the case of multiple PACS codes or genres associated with a paper or movie, only one code or genre is randomly selected. \textbf{H} is the layer-aggregated hypergraph.}
\label{tab:datasets}
\end{table*}

\section{Node properties}
\begin{figure}
    \centering  \includegraphics[width=\linewidth]{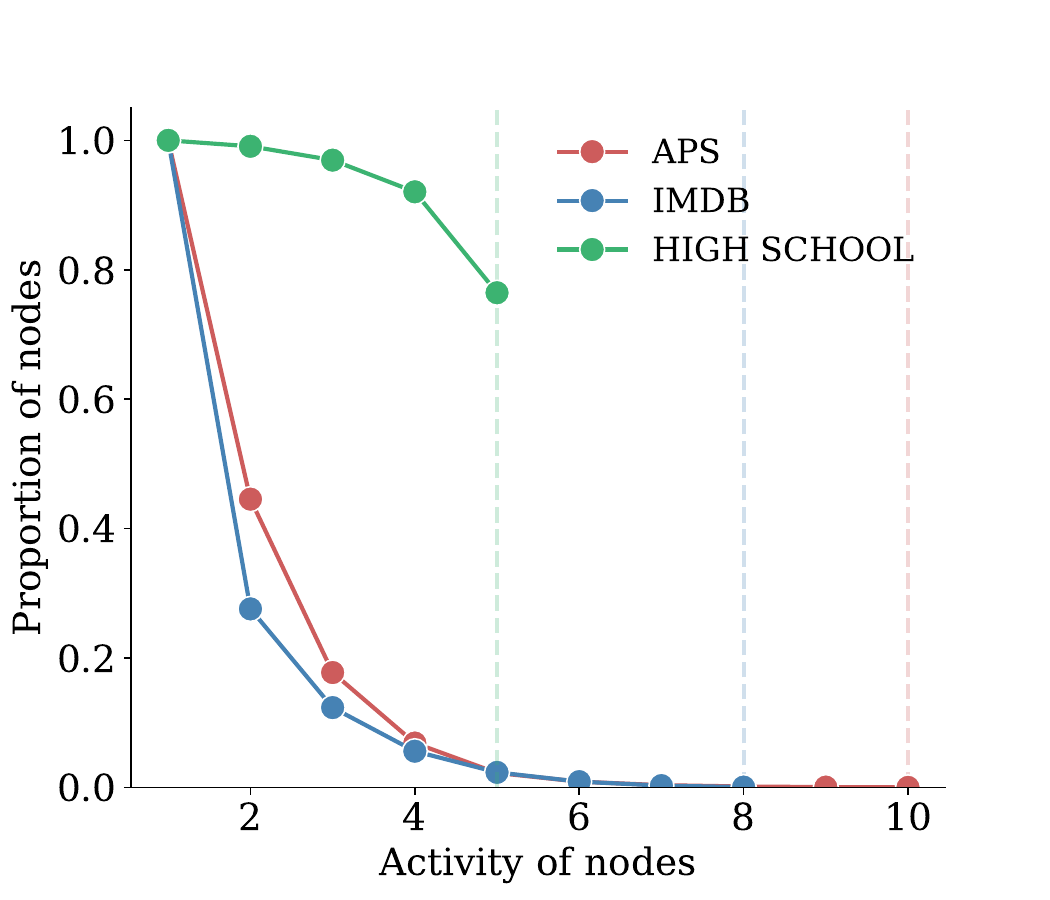}
    \caption{Proportion of nodes active in at least $x$ layers across three different datasets. Colored dashed lines indicate the number of layers in each respective dataset.}
    \label{fig:node_activity}
\end{figure}

We begin by investigating multiplex properties at the node level. The first basic measure we consider is \textit{node activity}~\cite{nicosia2015measuring}. A node $i$ is \textit{active} at layer $\alpha$ if $i$ participates in at least one interaction at layer $\alpha$. Fig.~\ref{fig:node_activity} shows statistics on nodes' simultaneous activity across multiple layers. Specifically, the  $y$-axis plots the proportion of nodes (from the total node count) active in at least $x$ layers. 
By definition, these curves exhibit a decreasing trend, with variations in the negative slopes reflecting the datasets' diversity. While it is uncommon for scientists and actors to be active across more than $2$ or $3$ layers, students tend to be active in all layers. In fact, the inactivity of a student in a specific layer implies their absence from school on that day. 

\begin{figure*}
    \centering
    \includegraphics[width=\linewidth]{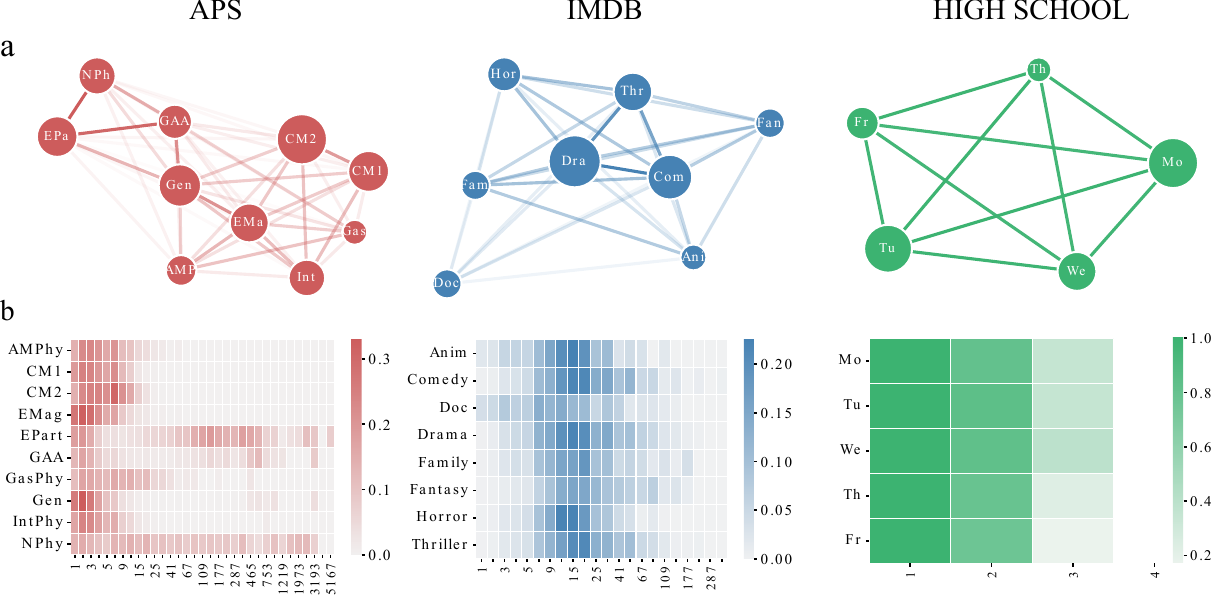}
    \caption{a) Each dataset is a graph in which vertices represent the layers of the multiplex hypergraphs and the thickness of an edge $(\alpha, \beta)$ quantifies the pairwise cosine similarity of layer activity matrices $\textbf{B}_\alpha, \textbf{B}_\beta$ associated with layers $\alpha$ and $\beta$. Vertex size is proportional to the number of nodes active in that layer. b) Matrix $L$ associated with each dataset. Rows are normalized by the number of nodes active in each layer. Interaction orders are binned exponentially.}
    \label{fig:nodes1}
\end{figure*}

So far, we have grouped all interactions together, regardless of their order. To obtain more detailed insights about higher-order interactions, we can examine node activity for each specific interaction order $d$.
To this scope, we introduce a list $\textbf{A}$ of \textit{node activity matrices}, one for each node $i$:
\begin{equation*}
    \textbf{A}_i = \{a_{\alpha d}\} = \begin{cases}
          1 & \text{if } i\text{ is active} \text{ at order } d \text{ in } \alpha \\
          0 & \text{otherwise}.
    \end{cases}
\end{equation*}

Similarly, one can define activity from a layer perspective and consider a list $\textbf{B}$ of \textit{layer activity matrices}, one for each layer $\alpha$:
\begin{equation*}
    \textbf{B}_\alpha = \{b_{i d}\} = \begin{cases}
          1 & \text{if } i\text{ is active} \text{ at order } d \text{ in } \alpha \\
          0 & \text{otherwise}.
    \end{cases}
\end{equation*}

 It can be useful to aggregate information about nodes and define an \textit{aggregated layer activity matrix} $L$ as: 
 \begin{equation*}
    L = \{l_{\alpha d}\} = |\{ \text{nodes active at order } d \text{ in } \alpha \}|
\end{equation*}


In Fig.~\ref{fig:nodes1}a, each dataset is represented as a graph where vertices are the layers of the multiplex hypergraph and links measure the similarity in activity patterns of two layers $\alpha$ and $\beta$, quantified as the cosine similarity of their node activity matrices $\textbf{B}_\alpha$ and $\textbf{B}_\beta$. The thicker the link, the higher the similarity. This figure emphasizes layers that not only share common active nodes, but also exhibit similar patterns of participation across different hyperedge orders. Particularly, a consistent higher-order similarity is observed across school days, reflecting recurring interaction patterns throughout the week. Other datasets show more heterogeneous behaviour, with documentary casts differing significantly from other layers, while drama and comedy casts exhibit similar patterns. 

Fig.~\ref{fig:nodes1}b shows the aggregated layer activity matrices $L$ for the three datasets. To account for variations in layer size, we normalize each row by the total number of active nodes in the respective layer. Distinct collaboration patterns emerge across the subfields of physics and movie genres. 
For instance, scientists in General or Electromagnetic Physics usually contribute to papers with a smaller number of co-authors, whereas co-author groups in fields like Elementary Particles and Nuclear Physics exhibit more variation in size. In movie collaborations, actor activity is concentrated in medium-size groups, typically between ten and twenty members. However, documentaries often feature smaller casts, while family and comedy movies tend to have larger ones. The figure once again highlights how students at school maintain a consistent group size in their interactions throughout the week. 

\begin{figure*}
    \centering
    \includegraphics[width=\linewidth]{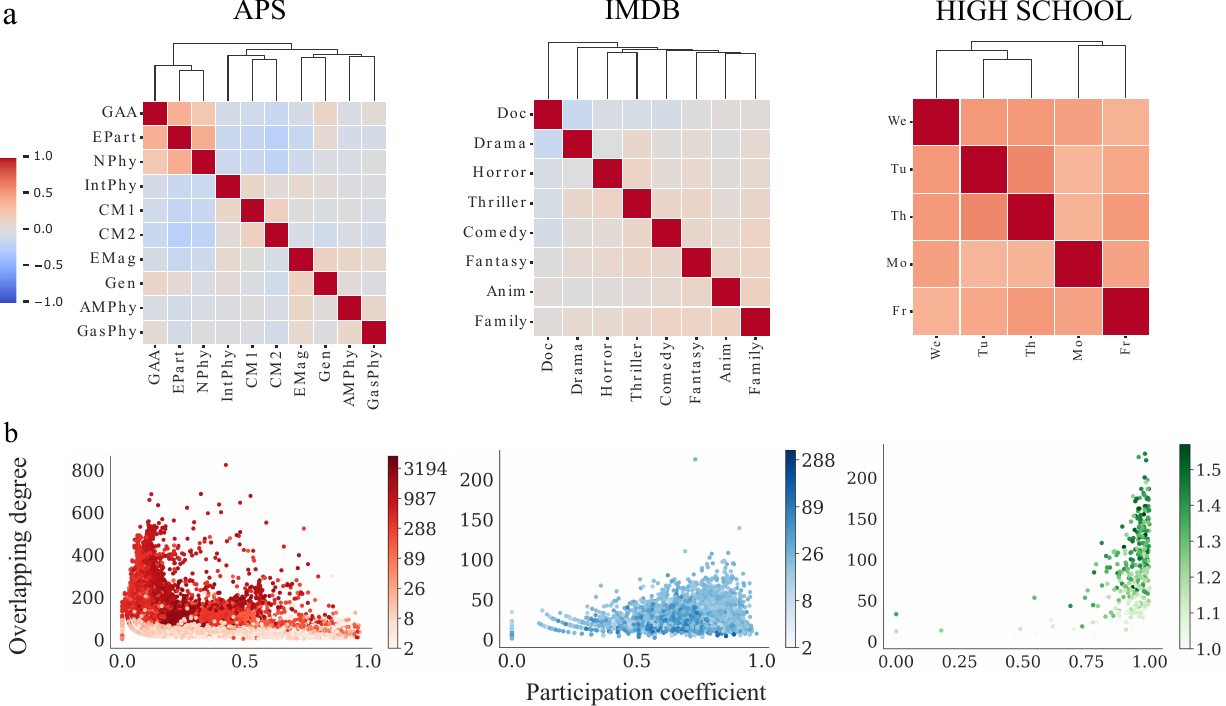}
    \caption{
    a) The heatmap shows the pairwise correlation between the degrees of nodes across different layers. The color scale indicates the strength of the correlation, with blue representing low correlation and red representing high correlation.
    b) A system unit \(i\) is represented as a point on a Cartesian plane, with the overlapping degree \(o_i\) on the \(y\)-axis, the participation coefficient \(P_i\) on the \(x\)-axis, and the average order of the interactions in which the unit is involved indicated by color intensity.}
    \label{fig:nodes2}
\end{figure*}

Similar to node activity, node degree (defined as the number of interactions in which a node participates) is another property that can be used to measure the activity across the different layers and interaction orders. We define a list $\textbf{K}$ of \textit{node degree matrices}, one for each node $i$:
\begin{align*}
    \textbf{K}_i &= \begin{aligned}[t]
        \{k_{\alpha d}\} &= |\{ \text{hyperedges of order } d \\
        &\quad \text{involving } i \text{ at } \alpha \}|
    \end{aligned}
\end{align*}

We use $k_{i \alpha}$ to denote the \textit{total number of interactions} involving $i$ in layer $\alpha$, irrespective of their order:
\begin{equation*}
    k_{i \alpha} = \sum_{d=1}^{D}{\textbf{K}_{i \alpha d}},
\end{equation*}
where $D$ is the order of the largest interaction in the dataset.

In Fig.~\ref{fig:nodes2}a, we analyze the correlation of node degree across layers, exploring the extent to which a node with a high or low degree in one layer similarly exhibits a high or low degree in another layer. The correlation matrix for physics collaborations uncovers a hierarchical structure, with strong correlations among specific subfields sharing commonalities and notable scientists, such as in Nuclear and Elementary Particles Physics. In contrast, the degree correlations among actors are generally weak, though certain genres, like thriller and horror, show similarities. A significant correlation in node degrees across consecutive days in \HS suggests stable and structured daily interaction patterns, implying that individuals with numerous interactions on one day tend to maintain similar levels of interactions on subsequent days, and vice versa.

We now define the \textit{overlapping degree} $o_i$ for a node $i$ as the total number of interactions involving $i$, irrespective of both layers or orders:
\begin{equation*}
    o_i = \sum_{\alpha=1}^M k_{i \alpha}
\end{equation*}

It can be interesting to measure (\textit{i}) how the overlapping degree of a node $i$ is spread across the layers, i.e., if the degree is concentrated in certain layers or if it is uniformly distributed; (\textit{ii}) how interactions involving node $i$ are spread across orders. We measure (\textit{i}) by defining the \textit{participation coefficient} $P_i$ of a node $i$ of the degree with respect to the layers:
\[
P_i = \frac{M}{M-1} \left[1 - \sum_{\alpha=1}^{M} \left(\frac{k_{i \alpha}}{o_i}\right)^2\right]
\]
where $k_{i\alpha}$ is the degree of node $i$ at layer $\alpha$, $o_i$ is the overlapping degree of node $i$ and $M$ is the total number of layers. We measure (\textit{ii}) by considering the average order of the interactions node $i$ participates in. 

In Fig.~\ref{fig:nodes2}b, we represent each unit $i$ of the different systems on a Cartesian plane, characterizing them across three distinct dimensions: their overlapping degree $o_i$ (on the $y$-axis), their participation coefficient $P_i$ (on the $x$-axis), and the average order of the interactions in which they are involved (indicated by color intensity). 
In general, such three dimensions provide different information about connectivity patterns and are only weakly correlated, or even uncorrelated.
In \APS, scientists are spread across the plane in terms of degrees and average interaction order, displaying an overall tendency towards specialization in a selected number of physics subfields, yet the behavior remains heterogeneous. \IMDB displays isolated outliers with a very high degree, low dispersion around the average interaction order, and a general tendency towards uniform participation across multiple genres. \HS shows students covering the entire spectrum of node degrees and average group orders, with an expected tendency to interact uniformly across school days.

\section{Hyperedge properties}

We now turn our attention to the properties of the hyperedges encoding interactions in the same three real-world systems.

\begin{figure*}
    \centering
    \includegraphics[width=\linewidth]{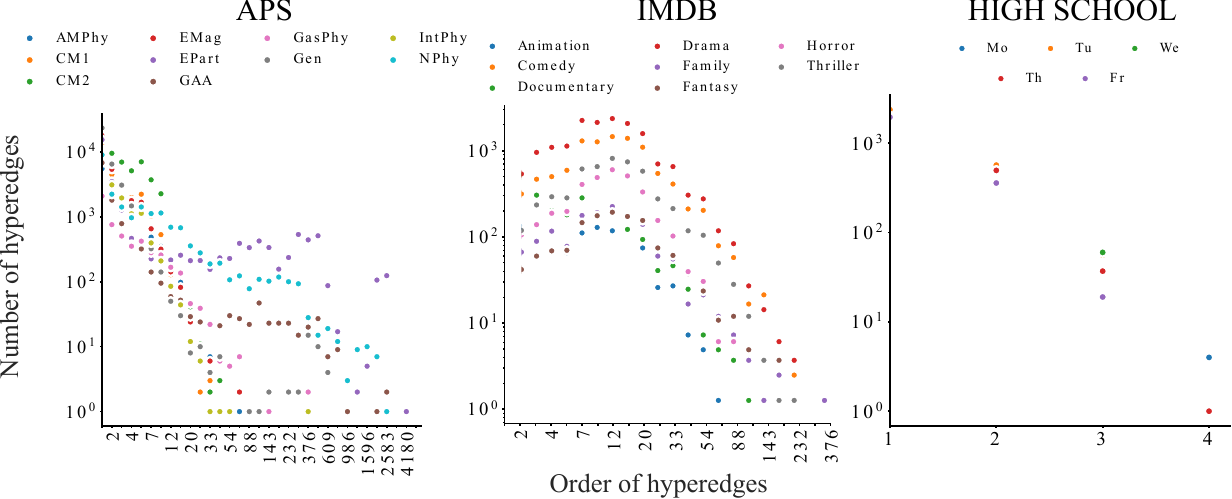}
    \caption{Distribution of hyperedge orders disaggregated by layers in each dataset. Colors distinguish between different layers, with interaction orders binned exponentially.}
    \label{fig:edges_dist}
\end{figure*}

We begin by considering the simplest measure for characterizing higher-order interactions, namely, the order of the groups. In Fig.~\ref{fig:edges_dist}, we plot the hyperedge order distribution disaggregated by layers. \APS and \IMDB reveal heterogeneity across layers, suggesting that different physics subfields and movie genres exhibit distinct patterns of collaboration in terms of the number of people involved in a paper or a movie cast. For example, genres such as documentaries and animated movies typically feature fewer actors compared to other genres. 
Conversely, papers in Elementary Particles and Nuclear Physics often include a larger number of authors compared to those in other areas of physics. The distributions in \HS are stable across layers, indicating that patterns of face-to-face interactions tend not to change over the days, with a general preference for smaller groups over larger ones.

\begin{figure}[b!]
    \centering
    \includegraphics[width=\linewidth]{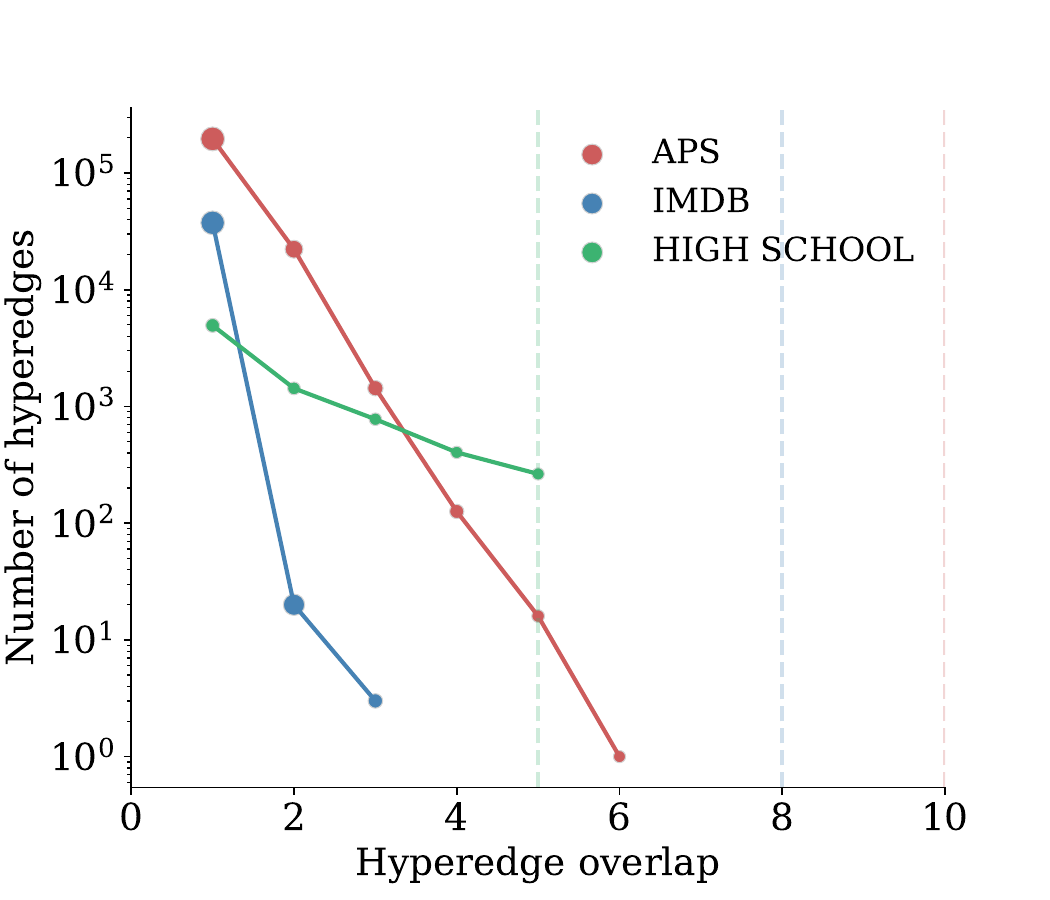}
    \caption{Number of hyperedges as a function of their overlap, i.e., the maximal number of layers in which an interaction repeats. Markers are scaled proportionally to the average order of hyperedges. Colored dashed lines indicate the corresponding number of layers in each dataset.}
    \label{fig:edge_overlap}
\end{figure}

Another property frequently studied in the context of multiplex networks is edge overlap, which measures the extent to which interactions among the same nodes tend to repeat across multiple layers. We define \textit{hyperedge overlap} as the maximum number of layers in which an interaction repeats exactly. In Fig.~\ref{fig:edge_overlap}, we present the distribution of hyperedge overlap, including information about the order of the interactions. As expected, \APS displays a high degree of hyperedge overlap, indicating that the same set of scientific authors consistently interact across multiple areas of physics. Conversely, for actors, hyperedge overlap decays very rapidly. Small interactions typically exhibit a higher degree of overlap than large interactions. Historically, edge overlap in multiplex networks with higher-order interactions has often been investigated by projecting hyperedges at different layers into cliques, frequently resulting in extremely high values of edge overlap. Our analysis suggests that patterns of hyperedge overlap are more complex and that projections of hyperedges can account for the high amount of overlap previously observed~\cite{battiston2014structural}.

\begin{figure*}
    \centering
    \includegraphics[width=\linewidth]{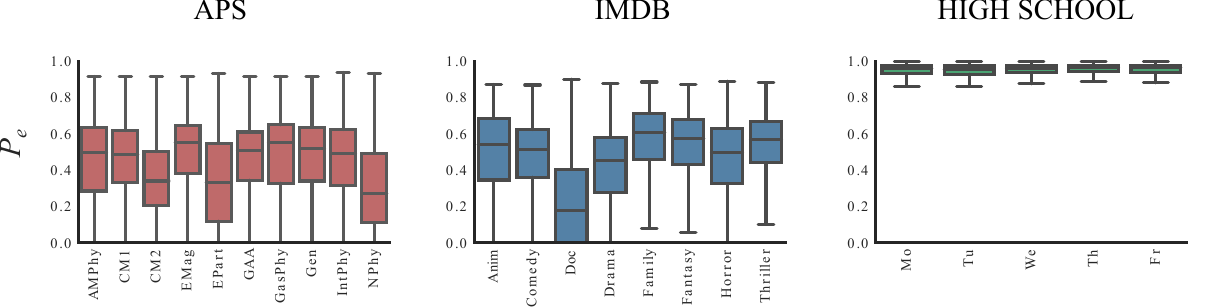}
    \caption{
    Boxplots showing the distribution of hyperedge participation coefficient $P_e$ across layers in each dataset.
    }
    \label{fig:edges_boxplot}
\end{figure*}

Finally, we assign a score $P_e$ to each hyperedge $e$, defined in terms of the participation coefficient of the nodes involved in the interaction:
\[
P_e = \frac{1}{|e|} \sum_{i \in e} P_{i}
\]
where $|e|$ represents the number of nodes participating in hyperedge $e$ and $P_{i}$ is the participation coefficient of node $i$, as defined in the previous section. 

This measure captures the tendency of hyperedges to connect nodes that either specialize in a few layers or act as generalists across multiple layers. In Fig.~\ref{fig:edges_boxplot}, we use boxplots to show distributions of $P$ for hyperedges in various layers. It is noteworthy that layers can display heterogeneous mean values for the participation coefficient of their hyperedges. For example, casts in documentaries and co-authors in Nuclear and Elementary Particles Physics tend to include specialists. On the other hand, family and thriller movies are more likely to feature generalist actors. In \HS, layers exhibit a consistent maximum mean value for the participation coefficient of hyperedges, attributed to students' regular attendance at school each day. 

\section{Mesoscale properties}
\begin{figure*}
    \centering
    \includegraphics[width=\linewidth]{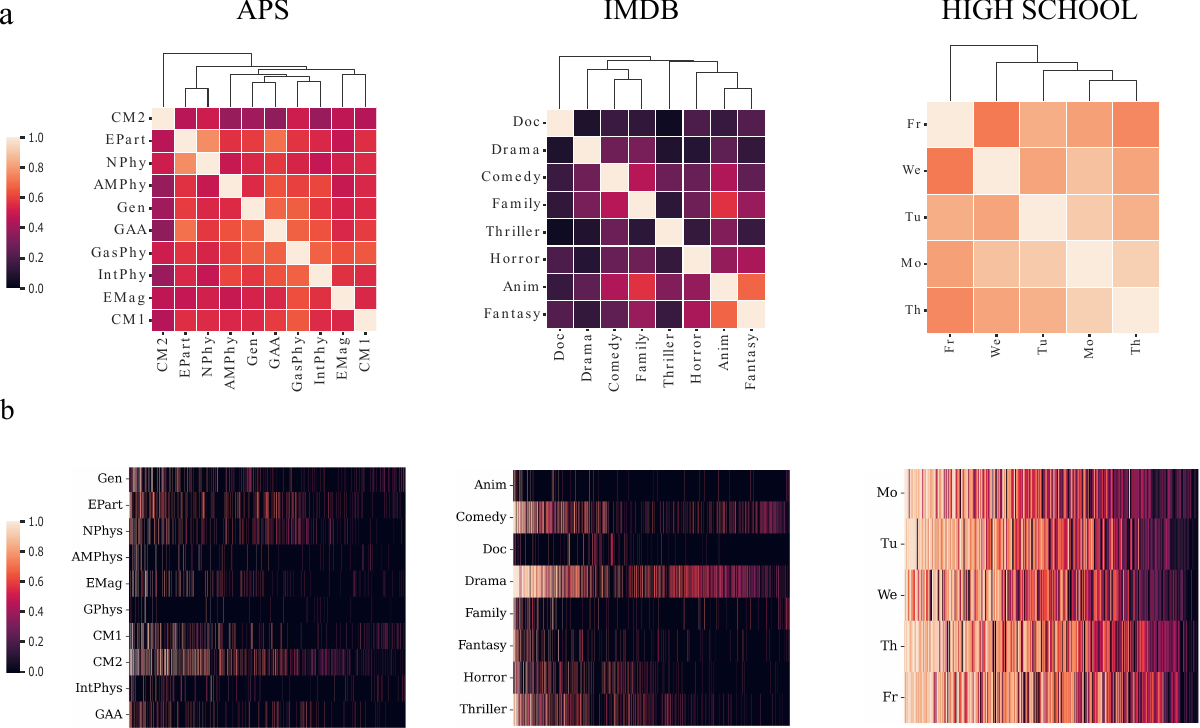}
    \caption{a) Heatmaps illustrating the similarity of community structures across layers, measured by Normalized Mutual Information (NMI), for the three datasets. In these heatmaps, high NMI values are represented by light colors, while low NMI values are represented by dark colors.} (b) Core-periphery score $c_i$ across layers for each node in the datasets, visualized as heatmaps. Rows are layers and columns represent nodes. Lighter colors are higher coreness values. Nodes are consistently sorted across layers based on their coreness in the aggregated hypergraphs, i.e., the hypergraph obtained by collapsing all layers into a single layer.
    \label{fig:mesoscale}
\end{figure*}

We finally shift our focus towards mesoscale structures, examining the emergence of communities and core-periphery structures within different layers of real-world hypergraphs. 

Communities are groups of nodes that display a higher degree of connectivity among themselves than with the rest of the nodes in the system. In hypergraphs, a \textit{community} is defined as a subset of nodes that tend to form cohesive units by participating in common hyperedges. When analyzing multiplex systems, it is typical to examine the similarities in community structures observed across various layers. In this direction, we employ a method for hard clustering, applied independently to each layer, which is an extension of the well-established Infomap algorithm to the case of hypergraphs~\cite{eriksson2021choosing}. In general, Infomap minimizes the map equation, which quantifies the description length required to represent the random walker's movements on the network~\cite{rosvall2009map}. This optimization effectively partitions the network into communities that best capture the inherent modular structure. The method is publicly available~\cite{eriksson2021choosing} and we have used default parameters.

To assess the similarity and consistency of community structure across different hypergraph layers, we use Normalized Mutual Information (NMI), taking into account the set of nodes active in both layers. NMI is defined as:

\[
\text{NMI}(C_1, C_2) = \frac{I(C_1, C_2)}{\sqrt{H(C_1) H(C_2)}}
\]

where \(I(C_1, C_2)\) denotes the mutual information between partitions \(C_1\) and \(C_2\). Mutual information quantifies the amount of information shared between the two partitions, i.e., measures how much knowing the community structure in one partition informs about the structure in the other. \(H(C_1)\) and \(H(C_2)\) represent the entropies of the partitions \(C_1\) and \(C_2\), respectively. By normalizing the mutual information \(I(C_1, C_2)\) with the geometric mean of these entropies, NMI adjusts for the variability in partition sizes and the number of communities, allowing for a fair comparison of community structures across different partitions. NMI ranges from \(0\) (indicating no mutual information) to \(1\) (indicating perfect agreement).

In Fig.~\ref{fig:mesoscale}a, we present the outcomes of this analysis through heatmaps, highlighting the strength of community structure similarities across layers. For example, community structures in animation and fantasy movie casts are closely related, as are those in comedy and family movie genres, whereas documentary casts show a completely uncorrelated structure. The community structure within \HS interactions remains consistent across days. Physics collaborations reveal a significant degree of similarity across fields, though some layers exhibit more pronounced similarities than others. Overall, communities tend to be preserved within physics subfields and school days, while movie genres often demonstrate predominantly uncorrelated communities.

We then direct our attention to core-periphery structures. Core-periphery structures delineate the existence of a group of central and tightly connected nodes, the core, surrounded by less densely connected peripheral nodes, forming a distinctive organizational pattern often crucial for system functionality~\cite{borgatti2000models}. 

We detect core-periphery structure for each layer independently using a method tailored for hypergraphs~\cite{tudisco2023core}. Such a method ranks nodes assigning to each node $i$ a value $c_i$, where $c_i$ is a real number within the range of $0$ to $1$. This value delineates the extent to which a node participates in the core (value closer to $1$) or the periphery (value closer to $0$) structure of the system. Vector $\mathbf{c}$ is selected for each layer $\alpha$ independently. Following the work by Tudisco and Higham~\cite{tudisco2023core}, for each layer $\alpha$ we select the vector $\mathbf{c}$ that optimizes the following function:

\[
\max_{\mathbf{c}} \, \sum_{e \in E_\alpha} \frac{1}{|e|} \cdot \left( \sum_{i \in e} c_i \right) \quad \text{subject to} \quad \| \mathbf{c} \|_2 = 1 \text{ and } c_i \geq 0 \text{ for all } i
\]

This continuous scale allows for a nuanced characterization of each node's role within the core-periphery framework. Additionally, by comparing the coreness score \( c_i \) for each node \( i \) across the different layers of a multiplex hypergraph, we can analyze the variation or consistency of a node centrality across the layers. This method is publicly available~\cite{tudisco2023core}. We use the implementation from Hypergraphx~\cite{lotito2023hypergraphx}.

To provide a visually appealing way of highlighting correlations of core-periphery structures and node behaviour across layers, Fig.~\ref{fig:mesoscale}b shows heatmaps in which rows are layers and columns are nodes, and each entry is coloured depending on node coreness. To compare the coreness value of single nodes across layers and visualize to which extent it keeps its core value, we maintain a consistent sorting of the nodes on the x-axis. For each dataset, nodes are sorted according to their core values in the aggregated hypergraphs (i.e., the hypergraph obtained by dropping information about layers and collapsing every hyperedge to a single layer). We observe that coreness values of nodes are maintained across layers exhibiting patterns similar to those seen in community structure correlations.

\section{Conclusions}

Networks have established themselves as a fundamental tool in a variety of disciplines to encode and study systems of interacting units. With the idea of capturing richer information about interactions, novel and more comprehensive network models have emerged:
(i) multiplex networks, describing links of different types, and (ii) hypergraphs, encoding non-dyadic ties.
 Bridging these two notions, in this work we have introduced a general set of measures to characterize the structure of multiplex hypergraphs at multiple scales. We introduced a description of nodes in terms of higher-order activity patterns and degrees, to quantify the extent and magnitude of node participation in interactions of different orders across layers. Nodes have been also characterized in terms of how their degree is correlated and spread in the different layers, and by their preferred order of group interactions. For hyperedges, we have studied their order distribution disaggregated by layers, highlighting different patterns of group interactions depending on the hyperedge type. We have quantified the extent to which hyperedge tends to repeat exactly in multiple layers and we have analyzed the layer-dependant property of hyperedges of linking nodes with low or high participation coefficients. Finally, we have analyzed hypergraphs at their mesoscale, quantifying similarities of communities and core-periphery participation across layers. We have validated our proposed measures on three datasets from different domains, describing collaboration patterns across physics subfields, movie genres and daily interactions among students. 

In summary, we believe that our measures can be useful in describing the structure of various empirical datasets characterized by both multiplex and higher-order interactions. We also hope that this initial characterization of multiplex hypergraphs will spark interest from a methodological perspective, such as proposing frameworks for extracting multiplex communities in hypergraphs. Further characterizations could be enhanced by considering the complex patterns of temporality in hyperedges, a common feature of higher-order systems~\cite{cencetti2021temporal, ceria2023temporal,gallo2023higher, iacopini2023temporal, digaetano2024percolation,arregui2024patterns,mancastroppa2024structural}.

\begin{widetext}

\section*{Code availability}
Multiplex measures for higher-order networks, as well as basic functions for handling multiplex hypergraphs, are available as part of Hypergraphx (HGX)~\cite{lotito2023hypergraphx}.

\section*{Data availability} 
Data is publicly available as part of HGX data repository~\cite{lotito2023hypergraphx}.

\section*{Acknowledgements}
F.B. acknowledges support from the Air Force Office of Scientific Research under award number FA8655-22-1-7025. A.M. acknowledges support from the European Union through Horizon Europe CLOUDSTARS project (101086248).

\end{widetext}

\bibliography{biblio}

\newpage
\begin{widetext}
\section*{Supplementary information}
\setcounter{figure}{0}
\setcounter{section}{0}
\renewcommand{\thefigure}{S\arabic{figure}}
\renewcommand{\thesection}{S\arabic{section}}

\section{Multiplex measures on randomized hypergraphs}
In this section, we investigate how our proposed multiplex measures for higher-order networks behave under re-wiring. This analysis helps in understanding the robustness of the observed network properties and the significance of the original structure.

We propose two different randomization methods. They are available as part of Hypergraphx~\cite{lotito2023hypergraphx}. Results have been averaged over ten realizations of the random models.
\newline

For this analysis we employ:

\begin{itemize}
    \item A \emph{configuration model} for hypergraphs~\cite{chodrow2020configuration} which preserves the degree sequences and the hyperedge order distributions of the original layers of the multiplex hypergraphs while randomizing the actual nodes involved in the hyperedges.
    \item A \emph{random model} for hypergraphs which preserves only the hyperedge order distributions of the original layers of the multiplex hypergraphs while randomizing the actual nodes involved in the hyperedges. 
\end{itemize}

By preserving the degree sequence in each layer, the configuration model maintains the node activity and the degree correlations across layers. Additionally, it preserves the distribution of overlapping degrees across different layers, and consequently, the node participation coefficient. However, by re-wiring interactions and thus randomizing the neighbors in each interaction, we disrupt the micro and mesoscale structure of the hypergraphs. To assess this, in Fig.~\ref{fig:si-persistence} we show the impact of the configuration model randomization process on the measure of hyperedge overlap in our empirical datasets. We notice a consistent drop in the number of overlapping hyperedges with respect to the original hypergraphs. In particular, after randomization, there are no hyperedges that appear in more than one layer in the case of movie collaborations.

\begin{figure}
    \centering
    \includegraphics[width=0.6\linewidth]{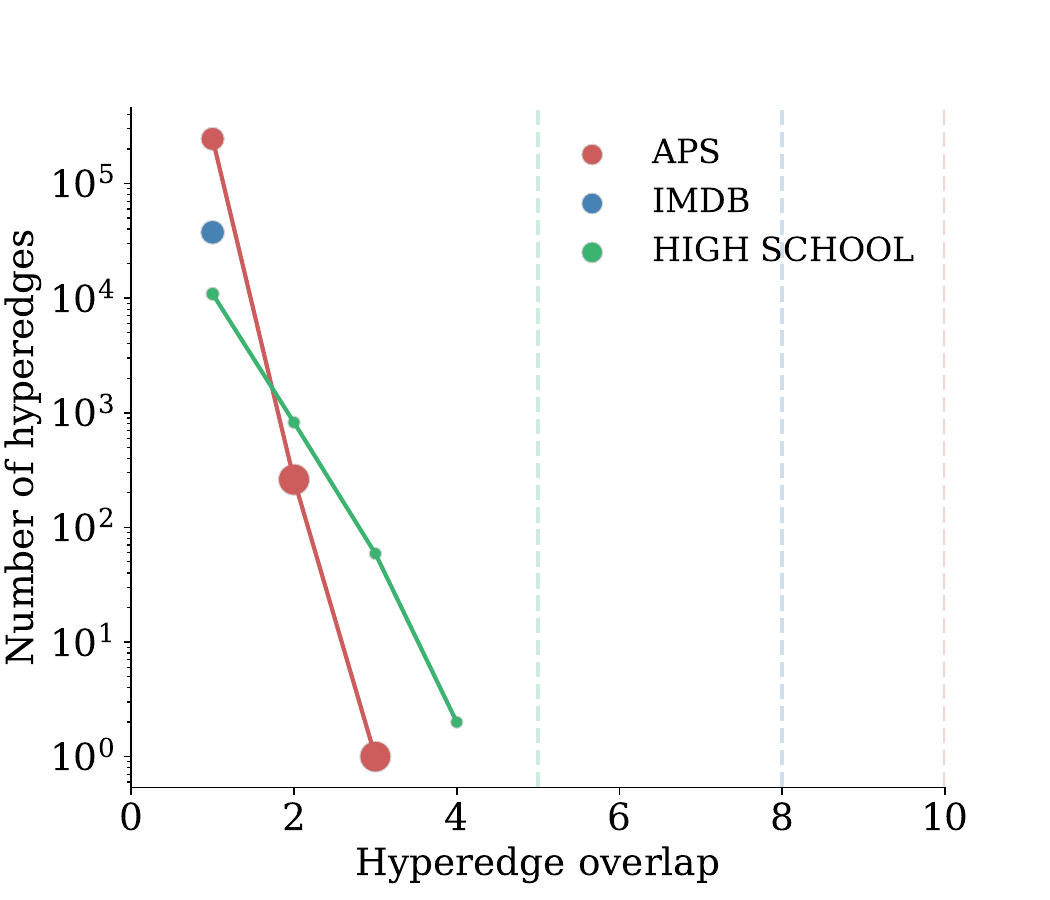}
    \caption{Impact of configuration model randomizations on hyperedge overlap. We notice a consistent decrease in overlapping hyperedges compared to the original data, with no hyperedges appearing in more than one layer for movie collaborations after randomization.}
    \label{fig:si-persistence}
\end{figure}

Additionally, in Fig.~\ref{fig:mesoscale_null} we show that the correlations between community structures are not preserved, as nodes tend to lose their preferential patterns of interactions under randomization. In particular, in \HS the algorithm collapses every node into a single community in every layer since the networks lack clear separations.

\begin{figure}
    \centering
    \includegraphics[width=\linewidth]{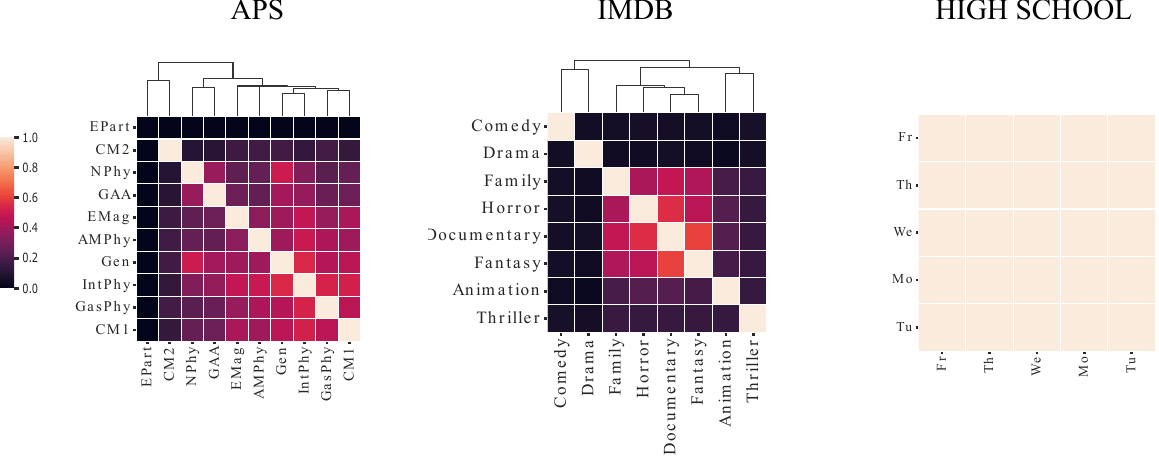}
    \caption{Configuration model randomization leads to the loss of preferential interaction patterns, with nodes losing correlated community affiliations across layers.}
    \label{fig:mesoscale_null}
\end{figure}

Our second random model is more disruptive, as it disintegrates structures to a greater extent. While this model preserves the hyperedge size distributions for each layer of a multiplex hypergraph, it randomizes the nodes involved in the interactions. This process is sufficient to dismantle node activity patterns and disrupt node degree correlations. Since node degrees are not preserved, node and hyperedge participation coefficients are also not maintained. It is important to notice that since all layers of a multiplex hypergraph share the same node set, the only property of a layer that is preserved is layer density. The rewiring of interactions allows more nodes to become active in new layers. This is shown in Fig.~\ref{fig:node_layers_activity_null}, in which we analyze nodes’ simultaneous activity across multiple layers and see an increase in such statistics with respect to original data.

\begin{figure}
    \centering
    \includegraphics[width=0.6\linewidth]{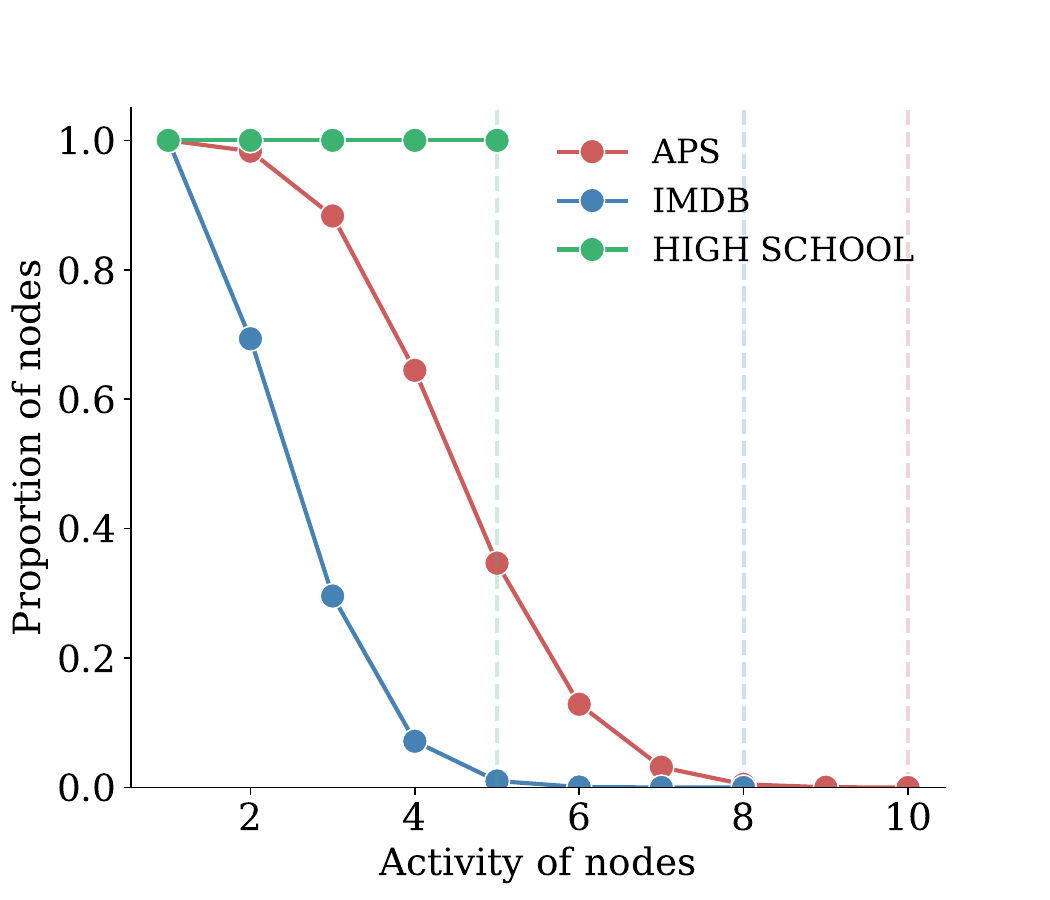}
    \caption{Rewiring preserves layer density but increases node activity across multiple layers. This is shown by the increase in nodes' simultaneous activity across layers compared to the original data.}
    \label{fig:node_layers_activity_null}
\end{figure}

The disruption of activity patterns is evident when computing the similarity among activity matrices between layers (Fig.~\ref{fig:node1_null}). In fact, the cosine similarity of the activity matrices drops to zero for both \APS and \IMDB, while \HS maintains a generally high level of similarity (explained by the abundance of interactions of size two and the high level of activity across layers). The disruption of activity patterns in \APS and \IMDB is caused by the rewiring of hyperedges with large cardinality, especially when these large hyperedges overlap in real-world data.

\begin{figure}
    \centering
    \includegraphics[width=\linewidth]{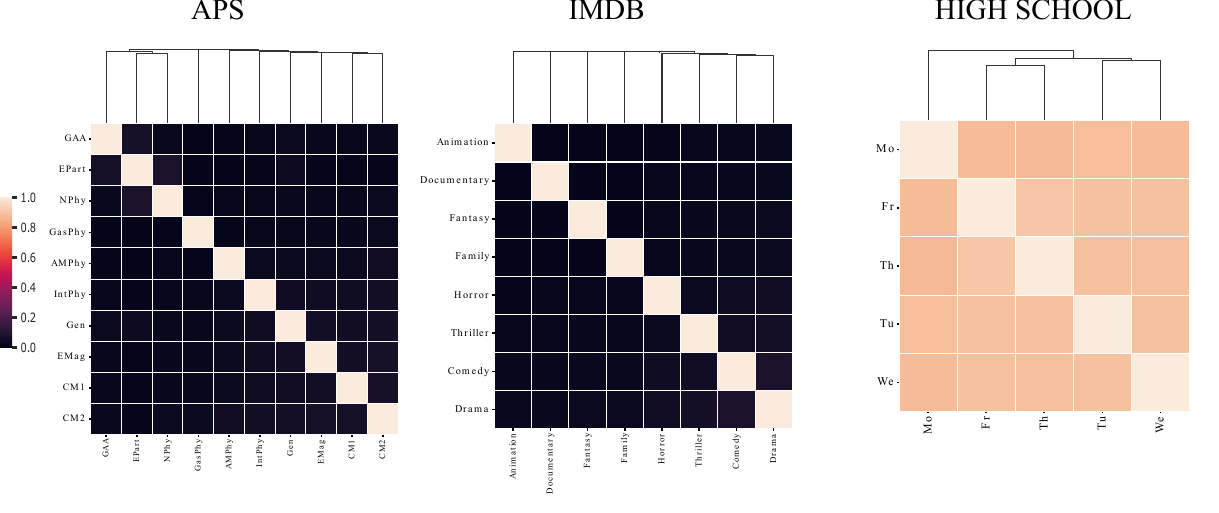}
    \caption{Similarity among activity matrices between layers drops to zero for \APS and \IMDB but remains high for \HS. This is due to the rewiring of large cardinality hyperedges, particularly when these hyperedges overlap in real-world data.}
    \label{fig:node1_null}
\end{figure}

Finally, it is important to examine the degree correlations across layers and the participation coefficient of nodes. The rewiring process significantly affects degree correlations, resulting in all pairs of layers being uncorrelated in terms of node degrees (Fig.~\ref{fig:node2_null}a). In Fig.~\ref{fig:node2_null}b we show that hubs tend to disappear (lower overlapping degrees than original data), average interaction size tends to decrease (due to overabundance of low order interaction in the data) and participation coefficient tends to increase (given by the increase in activity of nodes).

\begin{figure}
    \centering
    \includegraphics[width=\linewidth]{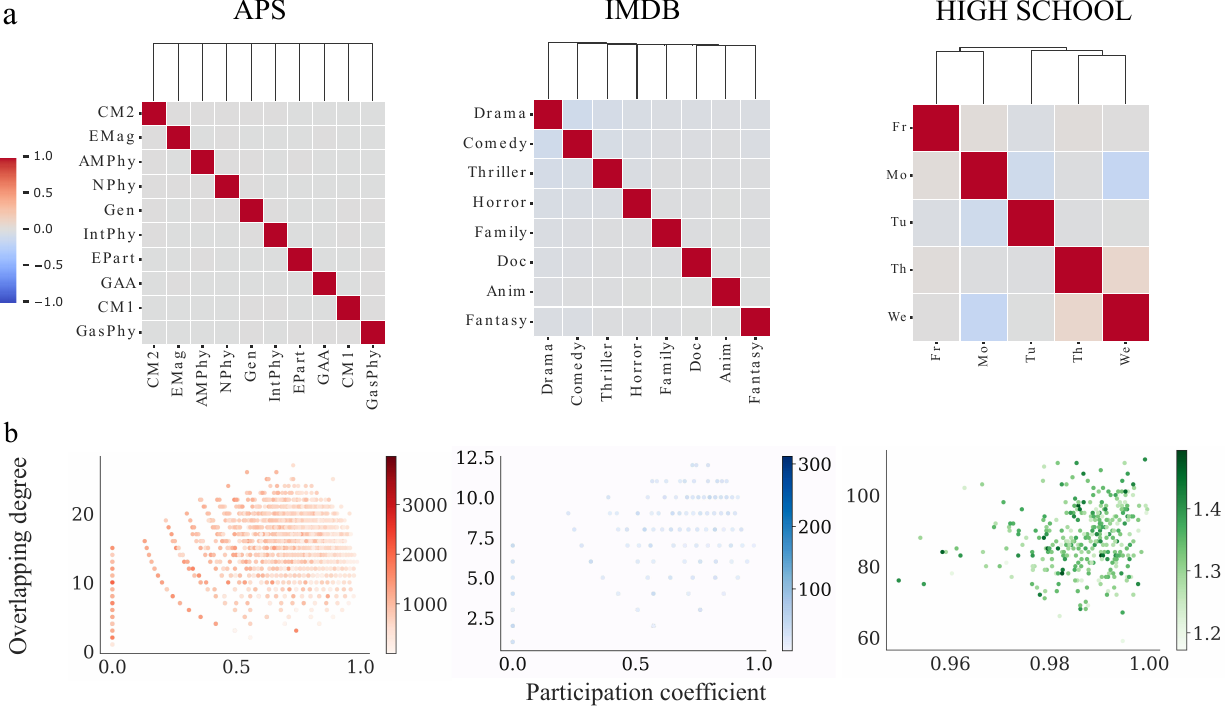}
    \caption{Effects of rewiring on multiplex hypergraphs properties. (a) Rewiring results in uncorrelated node degrees across all layer pairs. (b) Hubs disappear (lower overlapping degrees), average interaction size decreases, and participation coefficient increases due to higher node activity.}
    \label{fig:node2_null}
\end{figure}
\end{widetext}

\end{document}